\documentclass[12pt]{article}
\usepackage{amsmath,epsfig,graphicx,amssymb}
\usepackage{color,ulem}
\input{epsf} 
\usepackage{rotating}

\def\ltap{\raisebox{-.6ex}{\rlap{$\,\sim\,$}} \raisebox{.4ex}{$\,<\,$}} 
\def\gtap{\raisebox{-.6ex}{\rlap{$\,\sim\,$}} \raisebox{.4ex}{$\,>\,$}}

\newcommand\as{\alpha_{\mathrm{S}}} 
\newcommand\f[2]{\frac{#1}{#2}} 
\def\beq{\begin{equation}} 
\def\eeq{\end{equation}} 
\def\beeq{\begin{eqnarray}} 
\def\eeeq{\end{eqnarray}} 
 
\def\to{\rightarrow}

\begin{document} 
\begin{titlepage}
\begin{flushright}
ZU-TH 18/11\\
IFUM-984-FT
\end{flushright}

\renewcommand{\thefootnote}{\fnsymbol{footnote}}
\vspace*{2cm}

\begin{center}
{\Large \bf Diphoton production at hadron colliders:}
\vskip 0.15cm
{\Large \bf a fully-differential QCD calculation at NNLO }
\end{center}

\par \vspace{2mm}
\begin{center}
{\bf Stefano Catani${}^{(a)}$, Leandro Cieri${}^{(b)}$,
Daniel de Florian${}^{(b)}$,\\
Giancarlo Ferrera${}^{(c)}$}
and
{\bf Massimiliano Grazzini${}^{(d)}$\footnote{On leave of absence from INFN, Sezione di Firenze, Sesto Fiorentino, Florence, Italy.}}\\

\vspace{5mm}

${}^{(a)}$INFN, Sezione di Firenze and
Dipartimento di Fisica e Astronomia,\\ 
Universit\`a di Firenze,
I-50019 Sesto Fiorentino, Florence, Italy\\

${}^{(b)}$ 
Departamento de F\'\i sica and IFIBA, FCEYN, Universidad de Buenos Aires,\\
(1428) Pabell\'on 1 Ciudad Universitaria, Capital Federal, Argentina\\

${}^{(c)}$ 
Dipartimento di Fisica, Universit\`a di Milano and \\ INFN, Sezione di Milano, I-20133 Milan, Italy\\

${}^{(d)}$Institut f\"ur Theoretische Physik, Universit\"at Z\"urich, CH-8057 Z\"urich, Switzerland

\vspace{5mm}

\end{center}

\par \vspace{2mm}
\begin{center} {\large \bf Abstract} 

\end{center}
\begin{quote}
\pretolerance 10000

We consider direct diphoton
production in hadron collisions, and we compute the next-to-next-to-leading
order (NNLO) QCD radiative corrections at the fully-differential level.
Our calculation uses the $q_T$ subtraction formalism and it is implemented
in a parton level Monte Carlo program.
The program allows the user to apply arbitrary kinematical cuts on the
final-state photons and the associated jet activity, and to compute the
corresponding distributions in the form of bin histograms. We present selected
numerical results related to Higgs boson searches at the LHC and corresponding
results at the Tevatron.

\end{quote}

\vspace*{\fill}
\begin{flushleft}
October 2011

\end{flushleft}
\end{titlepage}

\setcounter{footnote}{1}
\renewcommand{\thefootnote}{\fnsymbol{footnote}}

Diphoton production is a relevant process
in hadron collider physics.
It is both a classical signal within the Standard Model (SM)
and an important background for Higgs and new-physics
searches.
The origin of the Electroweak symmetry breaking is currently
being
investigated at the LHC by searching for the Higgs boson and
eventually studying its properties. 
If the mass $m_H$ of the Higgs boson is low ($m_H \ltap 140$~GeV),
the preferred 
search mode at the LHC involves Higgs boson production via gluon fusion followed by the rare decay into a pair of photons.
Therefore, it is essential to count on an accurate theoretical description of
the various kinematical distributions associated to the production of pairs of
prompt photons with large invariant mass. Such task requires detailed computations of radiative corrections.

In this paper we are interested in the process $pp \rightarrow \gamma \gamma X$
(and the related process $p{\bar p} \rightarrow \gamma \gamma X$),
which, at the lowest order, occurs \textit{via} 
the quark annihilation subprocess $q\bar{q} \rightarrow \gamma \gamma$. The QCD corrections at 
the next-to-leading order (NLO) in the strong coupling $\as$ involve the  
quark annihilation channel and a new partonic channel, \textit{via} the
subprocess $qg \rightarrow \gamma \gamma q$. These corrections have been
computed and implemented in the fully-differential Monte Carlo codes
\texttt{DIPHOX} \cite{Binoth:1999qq}, \texttt{2gammaMC} \cite{Bern:2002jx} and 
\texttt{MCFM} \cite{Campbell:2011bn}. A calculation that includes the effects of
 transverse-momentum resummation is implemented in 
\texttt{RESBOS} 
\cite{Balazs:2007hr}.

At the next-to-next-to-leading order 
(NNLO), 
the $gg$ channel starts to contribute,
and
the large gluon--gluon luminosity makes this channel potentially sizeable. 
Part of the contribution from this channel,
the so called {\it box contribution}, was computed long ago \cite{Dicus:1987fk} and 
its size turns out to be comparable  
to the lowest-order result; for this reason,
the {\it box contribution} is customarily included in all the NLO computations of diphoton production.
The next-order gluonic corrections to the {\it box contribution}
(which are part of the N$^3$LO QCD corrections to diphoton production)
were computed in Ref.~\cite{Bern:2002jx} and found 
to have a moderate quantitative effect on the result of the `NLO+box' calculation.

Besides their {\it direct} production from the hard subprocess, photons can also
arise from fragmentation subprocesses of QCD partons. The computation of
fragmentation subprocesses requires (poorly known)
non-perturbative information, in the form of 
parton  
fragmentation functions of the photon.
The complete NLO single- and double-fragmentation contributions are implemented in \texttt{DIPHOX}
\cite{Binoth:1999qq}.
The effect of the fragmentation contributions 
is sizeably reduced by the photon isolation criteria that are 
necessarily
applied in hadron collider experiments to suppress the very large irreducible
background (e.g., photons that are faked by jets or produced by hadron decays). 
The standard cone isolation and the `smooth' cone isolation proposed
by Frixione \cite{Frixione:1998jh} are two of these criteria. The standard cone
isolation is easily implemented in experiments, but it only suppresses a fraction of the 
fragmentation contribution.
The smooth cone isolation (formally) eliminates the entire fragmentation 
contribution, but its experimental implementation is still in progress.

In this Letter we present the 
computation of the {\it full} NNLO QCD corrections
to direct diphoton production in hadron collisions.
We consider the inclusive hard-scattering reaction
\begin{equation}
\label{one}
h_1+h_2\to \gamma\gamma +X \;\;,
\end{equation}
where the collision of the two hadrons, $h_1$ and $h_2$,
produces the diphoton system 
$F \equiv \gamma\gamma$ with high invariant mass $M_{\gamma \gamma}$.
The evaluation of the
NNLO corrections to the process in Eq.~(\ref{one})
requires the knowledge of the corresponding partonic scattering amplitudes
with $X=2$~partons (at the tree level \cite{Barger:1989yd}), 
$X=1$~parton (up to the one-loop level \cite{Bern:1994fz})
and no additional parton (up to the two-loop level \cite{Anastasiou:2002zn})
in the final state.
The implementation of the separate scattering amplitudes in a complete
NNLO (numerical) calculation is severely complicated by 
the presence of infrared (IR) divergences that occur at intermediate stages. 
The $q_T$ subtraction formalism \cite{Catani:2007vq} is a method that handles
and cancels these unphysical IR divergences up to the NNLO.
The formalism applies to generic hadron collision processes that involve
hard-scattering production of a colourless high-mass system $F$.
 Within that framework \cite{Catani:2007vq}, the corresponding cross section is written as:
\begin{equation}
\label{main}
d{\sigma}^{F}_{(N)NLO}={\cal H}^{F}_{(N)NLO}\otimes d{\sigma}^{F}_{LO}
+\left[ d{\sigma}^{F+{\rm jets}}_{(N)LO}-
d{\sigma}^{CT}_{(N)LO}\right]\;\; ,
\end{equation}
where $d{\sigma}^{F+{\rm jets}}_{(N)LO}$ represents the cross section for the
production of the system $F$ plus jets at (N)LO accuracy\footnote{In the case of
diphoton production, the NLO calculation of 
$d{\sigma}^{\gamma\gamma+{\rm jets}}_{NLO}$ was performed in 
Ref.~\cite{DelDuca:2003uz}.}, and
$d{\sigma}^{CT}_{(N)LO}$ is a (IR subtraction) counterterm whose explicit expression \cite{Bozzi:2005wk}
is obtained from the resummation program of the logarithmically-enhanced
contributions to $q_T$ distributions. 
The combination of  $d{\sigma}^{F+{\rm jets}}_{(N)LO}$ and
$d{\sigma}^{CT}_{(N)LO}$ in Eq.~(\ref{main}) is numerically finite but the
two contributions separately diverge
in the limit of vanishing transverse momentum $q_T$
of the final-state system $F$.
In practice, we introduce a lower
limit on the transverse momentum $q_T$, such
that $q_T>q_{T{\rm cut}}$,
and we use a small finite value of $q_{T{\rm cut}}$.
Variations of $q_{T{\rm cut}}$ around this value allow us to estimate the systematic uncertainties related to the $q_T$ subtraction procedure.
The `coefficient' ${\cal H}^{F}_{(N)NLO}$, which also compensates for the subtraction
of $d{\sigma}^{CT}_{(N)LO}$,
corresponds to the (N)NLO truncation of the process-dependent perturbative function
\begin{equation}
{\cal H}^{F}=1+\f{\as}{\pi}\,
{\cal H}^{F(1)}+\left(\f{\as}{\pi}\right)^2
{\cal H}^{F(2)}+ \dots \;\;.
\end{equation}
The NLO calculation  of $d{\sigma}^{F}$ 
requires the knowledge
of ${\cal H}^{F(1)}$, and the NNLO calculation also requires ${\cal H}^{F(2)}$.

The general structure of ${\cal H}^{F(1)}$
is explicitly known \cite{deFlorian:2000pr}: 
${\cal H}^{F(1)}$ is directly obtained from the process-dependent scattering
amplitudes by using a process-independent relation.
Exploiting the explicit results of ${\cal H}^{F(2)}$ for Higgs
\cite{Catani:2007vq, Catani:2011kr} and vector boson \cite{Catani:2009sm} 
production,
we have generalized
the process-independent relation of 
Ref.~\cite{deFlorian:2000pr} to the calculation of the NNLO coefficient 
${\cal H}^{F(2)}$ (this general result is presented in a forthcoming paper).
Using this relation and the relevant scattering amplitudes
\cite{Barger:1989yd, Bern:1994fz, Anastasiou:2002zn}, we have explicitly
determined ${\cal H}^{F(2)}$ for diphoton production.

We have performed our fully-differential NNLO calculation of diphoton production
according to Eq.~(\ref{main}).
The NNLO computation is encoded
in a parton level
Monte Carlo program, in which
we can implement arbitrary IR safe cuts on the final-state
photons and the associated jet activity. 
The present formulation of the $q_T$ subtraction formalism \cite{Catani:2007vq}
is restricted to the production of colourless systems $F$ and, hence, it does not 
treat parton fragmentation subprocesses (here $F$ includes one or two coloured
partons that fragment).
Therefore, we concentrate on the direct production of diphotons, and 
we rely on the smooth cone isolation criterion \cite{Frixione:1998jh}.
Considering a cone of radius $r=\sqrt{(\Delta \eta)^2+(\Delta \phi)^2}$ around
each photon, we require that the total amount of hadronic (partonic) transverse energy $E_T$ 
inside the cone is smaller than $E_{T\, max}(r)$,
\begin{equation}
E_{T\, max}(r) \equiv  \epsilon_\gamma \,p_T^\gamma \left(\frac{1-\cos r}{1- \cos R}\right)^n \, ,
\end{equation}
where $p_T^\gamma$ is the photon transverse momentum; the isolation criterion
$E_T < E_{T\, max}(r)$ has to be fulfilled for all cones with $r\leq R$.
The isolation parameters
are set to the values $\epsilon_\gamma=0.5$, $n=1$ and $R=0.4$ in 
all the numerical results presented in this Letter.
We use the Martin-Stirling-Thorne-Watt (MSTW) 2008 \cite{Martin:2009iq} sets of parton distributions, with
densities and $\as$ evaluated at each corresponding order
(i.e., we use $(n+1)$-loop $\as$ at N$^n$LO, with $n=0,1,2$),
and we consider $N_f=5$ massless quarks/antiquarks and gluons in 
the initial state. The default
renormalization ($\mu_R$) and factorization ($\mu_F$) scales are set to the value
of the invariant mass of the diphoton system,
$\mu_R=\mu_F = M_{\gamma\gamma}$. The QED coupling constant $\alpha$ is fixed to $\alpha=1/137$.

We apply typical kinematical cuts \cite{ATLAS-CMS} that are
used by the ATLAS and CMS Collaborations in their Higgs search studies. We
require the harder photon to have a transverse momentum $p_T^{\rm harder}\geq
40$~GeV, while for the softer photon we demand $p_T^{\rm softer}\geq 25$~GeV.
The rapidity of both photons is restricted to $|y_\gamma| \leq 2.5$, and
the invariant mass of the diphoton system is constrained to lie in the range $20 \,{\rm GeV}\leq M_{\gamma\gamma} \leq 250\,{\rm GeV}$.

\begin{table}[htbp]
\begin{center}
\begin{tabular}{|c|c|c|c|}
\hline
$\sigma$ (fb)& LO & NLO & NNLO\\
\hline
\hline
$\mu_F=\mu_R=M_{\gamma\gamma}/2$ & $5045 \pm 1$ & $26581\pm 23$ & $42238 \pm 330$\\
\hline
$\mu_F=\mu_R=M_{\gamma\gamma}$ & $5712\pm 2$ & $ 26402\pm 25$ & $ 40269 \pm 250$ \\
\hline
$\mu_F=\mu_R=2M_{\gamma\gamma}$ & $6319\pm 2$ & $26045\pm 24$ & $ 38901 \pm 310$\\
\hline
\end{tabular}
\end{center}
\caption{{\it Cross sections for $pp\to \gamma\gamma+X$ at the LHC
($\sqrt{s}=14$~{\rm TeV}), including uncertainties of the numerical calculation.
The applied cuts are described in the text.}}
\label{tab:lhc}
\end{table}

\vspace{0.4cm}
\begin{figure}[htb]
\begin{center}
\begin{turn}{180}
\begin{tabular}{c}
\epsfxsize=10truecm
\epsffile{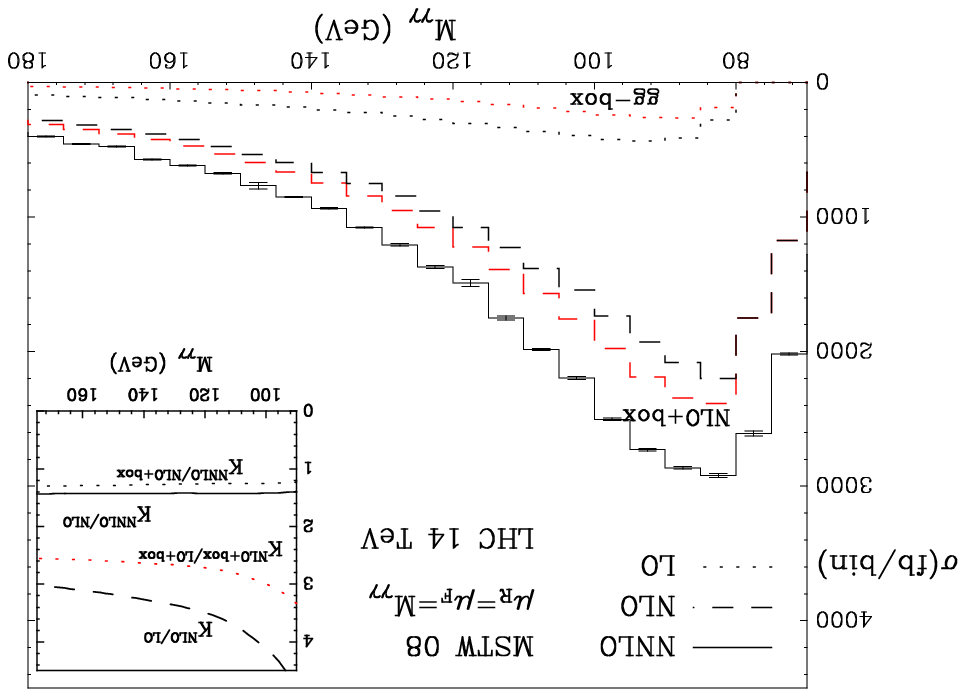}\\
\end{tabular}
\end{turn}
\end{center}
\caption{\label{fig:mass}
{\it Invariant mass distribution of the photon pair
at the LHC 
($\sqrt{s}=14$~{\rm TeV}): LO (dots), NLO (dashes) and NNLO (solid) results
with corresponding statistical uncertainties of the Monte Carlo integration. We
also present the results of the box and NLO+box contributions. The inset plot
shows the corresponding {\rm K}-factors.}}
\end{figure}

We start the presentation of our results by considering diphoton production
at the LHC ($\sqrt{s}=14$~TeV). In Table~\ref{tab:lhc}, we report the results
of the accepted cross section at LO, NLO and NNLO.
We have fixed $\mu_F=\mu_R=\mu$ and we have considered three values of 
$\mu/M_{\gamma\gamma}$ ($\mu/M_{\gamma\gamma}=1/2,1,2$).
As for the subtraction procedure, we have used $q_{T{\rm cut}}=0.04$ GeV.
The numerical errors at LO and NLO estimate the statistical uncertainty of the Monte Carlo integration. At NNLO they also include an estimate of the systematic uncertainty from using a finite $q_{T{\rm cut}}$.

\begin{figure}[htb]
\begin{center}
\begin{turn}{180}
\begin{tabular}{c}
\epsfxsize=10truecm
\epsffile{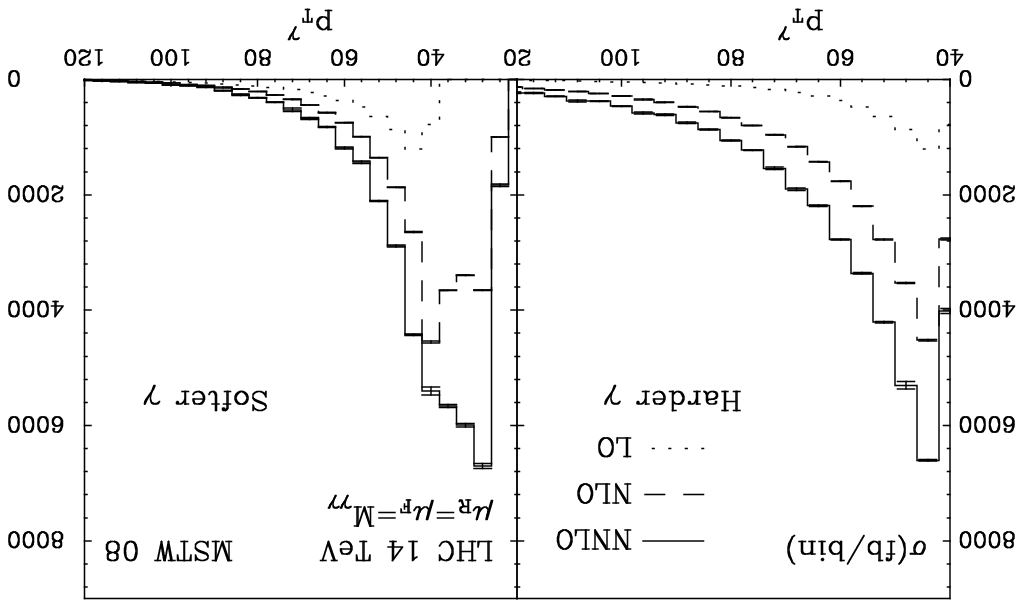}\\
\end{tabular}
\end{turn}
\end{center}
\caption{\label{fig:dist}
{\it Transverse-momentum distribution of the harder (left)  and softer (right)
photon at the LHC (notation as in Fig.~\ref{fig:mass}).}}
\end{figure}

We note that 
the value of the cross section remarkably increases with the perturbative order
of the calculation. This increase is mostly due to the use of {\it very
asymmetric} (unbalanced) cuts on the photon transverse momenta. At the LO,
kinematics implies that the two photons are produced with equal transverse
momentum and, thus, both photons should have $p_T^{\gamma}\geq 40$~GeV. 
At higher orders, the final-state radiation of additional partons opens a new
region of the phase space, 
where $40$~GeV $\geq p_T^{\rm softer}\geq 25$~GeV. Since photons can copiously be
produced with small transverse momentum (see also Fig.~\ref{fig:dist} and the
related discussion), the cross section receives a sizeable
contribution from the enlarged phase space region. This effect is further
enhanced by the opening of a new large-luminosity partonic channel at each
subsequent perturbative 
order. For example, at NLO the $qg$ channel accounts for about 80\% of the
increase of the cross section.
Therefore, it is not unexpected that a naive analysis of scale dependence (as
presented in Table~\ref{tab:lhc}) 
underestimates
the size of the higher-order corrections.

\vspace{0.4cm}

We now consider few kinematical distributions. The corresponding results are obtained with $q_{T{\rm cut}}=0.1$ GeV. 
In Figure \ref{fig:mass} we compare the LO, NLO and NNLO invariant mass
distributions at the default scales. We also plot the gluonic {\it box
contribution} (computed with NNLO parton distributions) and its sum with the full NLO result.
 The inset plot shows the K-factors defined by 
the ratio of the cross sections at two subsequent perturbative orders.
We note that ${\rm K}^{NNLO/NLO}$ is sensibly smaller than ${\rm K}^{NLO/LO}$,
and this fact indicates 
an improvement in the convergence of the perturbative expansion. 
In particular,
the impact of the full NNLO corrections turns out to be reasonably moderate, 
with a K-factor, defined as the ratio between the NNLO and NLO+box distributions,
of about ${\rm K}\simeq 1.25$. 
We find that about 13\% of the NNLO corrections is due to
the total contribution of the $gg$ channel
(which is roughly two thirds of the {\it box contribution}),
while almost
60\% still arises from the next-order corrections to the $qg$ channel.
The NNLO calculation includes the perturbative corrections from the entire phase
space region 
(in particular, the next-order correction to the dominant $qg$ channel)
and the contributions from all possible partonic channels (in particular,
a fully-consistent treatment of the {\it box contribution} to 
the $gg$ channel\footnote{The calculation \cite{Bern:2002jx} of the next-order
gluonic corrections to the  {\it box contribution} indicates an increase of 
the NNLO result by less than 10\% if $M_{\gamma\gamma} \gtap 100$~GeV.}).
Owing to these reasons, the NNLO result can be considered a reliable estimate of
direct diphoton production, although further studies (including independent
variations of $\mu_R$ and $\mu_F$, and analyses of kinematical distributions)
are necessary to quantify the NNLO theoretical uncertainty.

In Fig.~\ref{fig:dist} we  show results on more exclusive observables:
the $p_T$ distributions of the harder (left-hand plot) and softer
(right-hand plot) photon.
The statistical errors of the Monte Carlo integration are at the per-cent level
and hardly visible in Fig.~\ref{fig:dist}.
As previously anticipated, in the right-hand plot
we observe that significant NLO and NNLO contributions to the cross section 
originate from the phase space region (25~GeV $\leq p_T^{\rm softer} \leq 40$~GeV)
that, at LO, is kinematically forbidden by the asymmetric transverse-momentum
cuts. 
In this low-$p_T$ region, the production mechanism of the softer photon is
dynamically enhanced (the production probability is roughly
proportional to
$\as \times (p_T^{\rm harder}/p_T^{\rm softer}) \ln(p_T^{\rm harder}/p_T^{\rm
softer})$,
if $p_T^{\rm softer}/p_T^{\rm harder} \sim (p_T^{\rm softer}/M_{\gamma
\gamma})^2 \ll 1$),
and this is responsible for a substantial part of the large higher-order
corrections observed in Table~\ref{tab:lhc}, Fig.~\ref{fig:mass} and also in the 
$p_T$ distribution of the harder photon.
Comparing the $p_T$ distributions of the harder and softer photon in the 
high-$p_T$ region ($p_T \gtrsim 50$~GeV), we also observe that the 
distribution of the softer photon receives higher-order corrections that are
sensibly smaller. This decrease is expected: if both photons have high $p_T$,
the effect of the very asymmetric transverse-momentum cuts is reduced.

We also comment on the $p_T$ distribution of the softer photon in the region
around the LO threshold ($p_T^{\rm softer} \sim 40$~GeV). Here the LO result has
a step-like behaviour, and this necessarily produces \cite{Catani:1997xc}
integrable logarithmic singularities at each subsequent perturbative order.
The peak of the NLO distribution at $p_T^{\rm softer} \sim 40$~GeV
is an artifact of these perturbative instabilities. The instability is cured by
all-order perturbative resummation, which eventually leads to a smooth
$p_T$ distribution with a shoulder-like behaviour \cite{Catani:1997xc} 
in the vicinity of 
the LO threshold.
This physical behaviour can be approximated
(mimicked) in the NLO and NNLO calculations by smearing the distribution over a
bin (with a sufficiently-large size) centered around 
$p_T^{\rm softer} \sim 40$~GeV.

\vspace{0.4cm}

\begin{figure}[htb]
\begin{center}
\begin{turn}{180}
\begin{tabular}{c}
\epsfxsize=10truecm
\epsffile{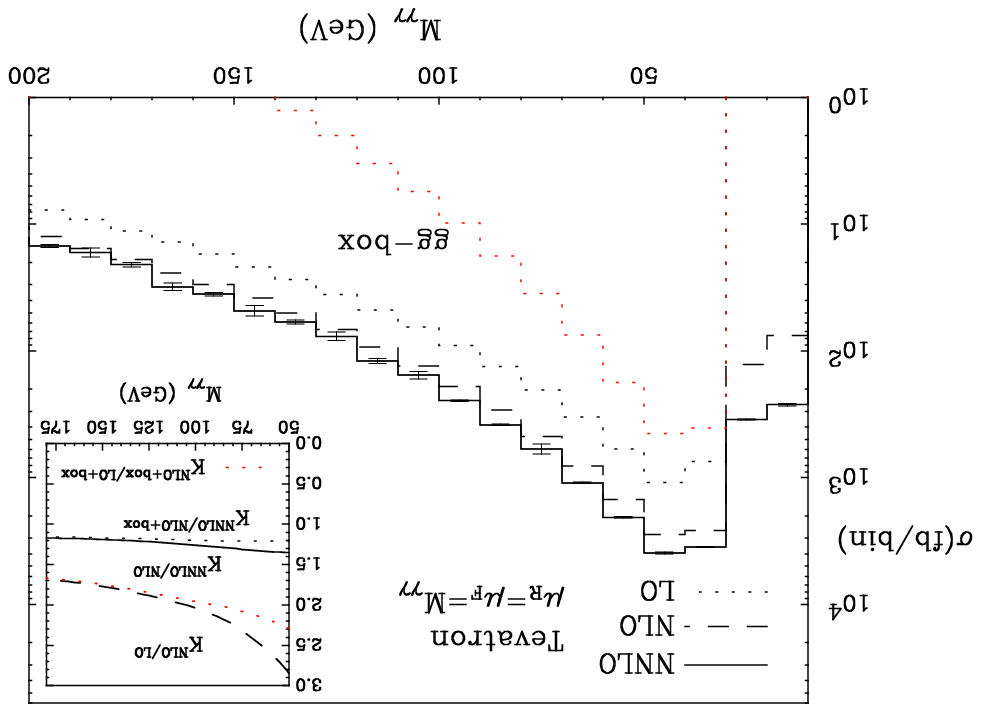}\\
\end{tabular}
\end{turn}
\end{center}
\caption{\label{fig:tev}
{\it 
Same as Fig.~\ref{fig:mass}, but for Tevatron kinematics 
(as described in the text).
}}
\end{figure}
In Figure~\ref{fig:tev}, we present the invariant mass distribution for 
diphoton production at the Tevatron ($\sqrt{s}=1.96$~TeV). We require  the
harder and softer photon to have a transverse momentum larger than  17~GeV and
15~GeV, respectively. The rapidity of both photons is restricted to $|y_\gamma| \leq 1$.
We note that the increase from the LO to the NLO result is considerably smaller than 
in Fig.~\ref{fig:mass}: this is mostly due to the use of photon
transverse-momentum cuts that are only slightly asymmetric.
In the region where $ M_{\gamma\gamma} \gtrsim 80$~GeV, 
the relative impact of the {\it box contribution} is smaller than at the LHC:
this is a consequence of the higher values of parton momentum fractions, $x$,
that are probed by Tevatron kinematics.
Nevertheless, the NNLO corrections 
(which are dominated by the next-order correction to the $qg$ channel)
still increase the result at the previous order by roughly 20\%.

We have presented the  
calculation of the cross section for diphoton production
up to the complete NNLO in QCD perturbation theory.
At the NNLO, all the contributions from the $gg$ channel are included in a
fully-consistent (and unambiguous) manner.
Considering the illustrative isolation and kinematical cuts implemented in this
paper, we find increasing effects of about 20\%$-$30\%
with respect to computations at the previous perturbative order.
Our calculation is
directly implemented in a parton level Monte Carlo program.
This feature makes it particularly suitable for practical applications
to the computation of distributions in the form of bin histograms.
A public version of our program is available at {\tt http://www.physik.uzh.ch/\textasciitilde lcieri/}.

\noindent {\bf Acknowledgements.}
This work was 
supported in part by UBACYT, CONICET, ANPCyT, INFN and the Research Executive Agency (REA) of the European Union under the Grant Agreement number PITN-GA-2010-264564 (LHCPhenoNet). We thank the Galileo Galilei Institute 
for Theoretical Physics
for the hospitality during the completion of this work.

\end{document}